\documentclass[floatfix,superscriptaddress,twocolumn,showpacs,prb,amsmath,amssymb]{revtex4} 

\usepackage{graphicx} 
\usepackage{amsmath}

\newcommand{\apo}{Ag$_5$Pb$_2$O$_6$}
\newcommand{\sub}[1]{$_{\mathrm {#1}}$}
\newcommand{\subm}[1]{_{\mathrm {#1}}}

\newcommand{\degc}{$^{\circ}$C}
\newcommand{\Tc}{T\subm{c}}
\newcommand{\Hc}{H\subm{c}}
\newcommand{\Hsc}{H\subm{sc}}
\newcommand{\dash}{^{\prime}}

\begin{document}

\title{Exceptional type-I superconductivity of the layered silver oxide Ag$_5$Pb$_2$O$_6$}

\author{Shingo~Yonezawa}
\affiliation{Department of Physics, Graduate School of Science, 
Kyoto University, Kyoto 606-8502, Japan}

\email{yonezawa@scphys.kyoto-u.ac.jp}

\author{Yoshiteru~Maeno}
\affiliation{Department of Physics, Graduate School of Science, 
Kyoto University, Kyoto 606-8502, Japan}
\affiliation{International Innovation Center, 
Kyoto University, Kyoto 606-8501, Japan}

\date{\today}


\begin{abstract}
We report zero-resistivity transition and the details of magnetic transition of 
a layered silver oxide Ag$_5$Pb$_2$O$_6$ single crystal,
which make definitive evidence of superconductivity in this compound.
In the AC susceptibility of a mono-crystal, 
we observed large supercooling, as well as positive peaks in the real part of the susceptibility 
indicating the reversibility of magnetic process.
These observations reveal that Ag$_5$Pb$_2$O$_6$ is probably the first
oxide that shows type-I superconductivity.
Evaluation of the superconducting parameters 
not only gives confirming evidence of type-I superconductivity, 
but also indicates that it is a dirty-limit superconductor.
We also analyze supercooling to determine the upper limit of the Ginzburg-Landau parameter.
\end{abstract}


\maketitle


In the last two decades, research of oxide superconductors is one of the most actively-studied fields
in solid state physics~\cite{Nagata1999, Cava2000}. 
Copper oxide high-$\Tc$ superconductors~\cite{Bednorz1986} discovered in 1987
made the greatest impact to the field.
Sr\sub{2}RuO\sub{4}~\cite{Maeno1994Nature}, 
with accumulating evidence for a spin-triplet superconductor~\cite{Mackenzie2003RMP},
has attracted much attention.
More recently, Na$_x$CoO\sub{2}$\cdot y$H\sub{2}O with a triangular lattice~\cite{Takada2003}
is widely studied because of the coexistence of superconductivity and geometrical frustration
and of possible novel superconducting phenomena.
We note here that the unconventional superconductors of oxides listed above have layered structures, 
and that it is believed a quasi-two-dimensional crystal structure is favorable for unconventional superconductivity.
As possible candidates for novel unconventional superconductivity,
silver oxides are particularly worth investigation,
since they might have electronic structures analogous to the high-$\Tc$ cuprates.
However, the only silver oxide superconductors reported so far were
cubic clathrate salts Ag\sub{7}O\sub{8}$X$ ($X$=NO\sub{3}, HF\sub{2}, etc.)~\cite{Robin1966} found in 1966.
Curiously, 
no other silver oxide superconductors have been reported for nearly 40 years,
let alone those with layered structures.

Here we report the discovery of superconductivity in \apo,
with $\Tc$ of 52.4~mK,
an eagerly-awaited and the very first layered silver oxide superconductor.
What is more, we clarified that \apo\ is a type-I superconductor.
This fact is rather surprising, 
since most type-I superconductors are pure metals and
only a handful are known among compounds and alloys.
To the best of our knowledge, reported compound type-I superconductors are only
ZrB\sub{12}~\cite{Tsindlekht2004}, 
YbSb\sub{2}~\cite{Yamaguchi1987}, 
LaPd\sub{2}Ge\sub{2}~\cite{Hull1981}, 
$M$Pd\sub{2}Si\sub{2} ($M$=Lu, Y, La)~\cite{Palstra1986}, 
TaSi\sub{2}~\cite{Gottlieb1992}, 
AuIn\sub{2}~\cite{Soulen1977},
C$_x$K~\cite{Kobayashi1981} (intercalation),
LaRh\sub{2}Si\sub{2}~\cite{Palstra1986};
thus \apo\ is the first oxide type-I superconductors.

\begin{figure}
\includegraphics[width=0.44\textwidth]{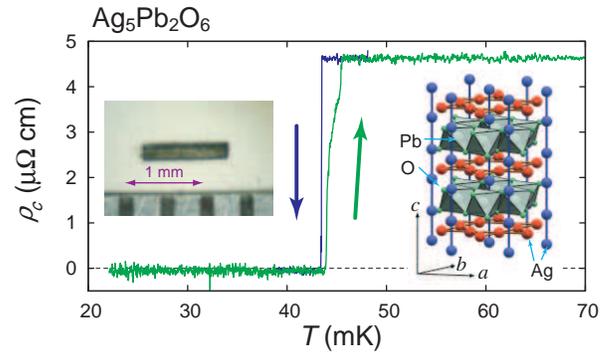}
\caption{(color online)
Temperature dependence of the out-of-plane resistivity $\rho_c$ of Ag$_5$Pb$_2$O$_6$ below 70~mK.
The sweep rate was approximately 0.05~mK/min.
Hysteretic behavior at the transition is attributable to a residual magnetic field.
The inset photo on the left shows the single crystal used for the measurements.
The inset on the right shows the crystal structure of Ag$_5$Pb$_2$O$_6$. Red and blue spheres
represent the silvers on the Kagome lattice and the chain, respectively.
\label{fig:zero-resistivity}}
\end{figure}

\apo, which was first reported by Bystr\"{o}m and Evers in 1950~\cite{Bystrom1950},
has a rather interesting crystal structure (see the inset of Fig.~\ref{fig:zero-resistivity})
consisting of a silver Kagome lattice parallel to its $ab$ plane 
and silver chains along the $c$ axis~\cite{Jansen1990}.
This silver oxide exhibits metallic conductivity.
Band calculation by Brennan and Burdett~\cite{Brennan1993} 
shows that its conductivity mainly comes from Ag5s orbital,
and that its Fermi surface has a quasi-three-dimensional character
because both the silver chain and Kagome lattice 
contribute to the density of states at the Fermi level.
Interestingly, the resistivity behaves as $\rho=AT^2+\rho_0$ in an unusually wide range of temperature,
down to below 4~K and up to room temperature~\cite{Yonezawa2004PRB}.
This means that unknown strong scattering mechanism dominates over the usual electron-phonon scattering.
Superconductivity of \apo\ was recently suggested by the present authors~\cite{Yonezawa2004PRB}.
We reported quite a large diamagnetic signal
in the AC susceptibility measured using a cluster of single crystals below 48~mK
but could not obtain zero resistivity at that time.
We finally observed zero resistivity by improving experimental techniques,
and present in this paper not only the observation but also the details of superconducting properties of \apo\ 
for the first time.


In the experiments, we used single crystals  
of \apo\ grown by the self-flux method, 
from mixture of 5-mmol AgNO$_3$ and 1-mmol Pb(NO$_3$)$_2$~\cite{Yonezawa2004PRB}.
All the measurements reported here were performed with a ${}^4$He-${}^3$He dilution refrigerator
(Cryoconcept, Model DR-JT-S-100-10), covering the measurement temperatures as low as 16~mK.
The resistivity
was measured using a conventional four-probe method with an AC current of 
10.4~$\mu$A rms at 163~Hz with a hexagonal-stick single crystal 
which fits in $0.14\times 0.21\times 1.15\ \mathrm{mm}^3$.
We used pure gallium to attach electrical wires of copper to the sample crystals.
We note here that one must keep the temperature of the electrodes 
well below the melting point of gallium (29\degc) 
all the time after soldering in order to avoid the electrical contacts getting worse.
We avoided using gold wires
because gallium easily dissolves gold.
The AC susceptibility was measured by a mutual inductance method.
We fabricated a very small and highly-sensitive cell
by winding a 50-$\mu$m-diameter copper wire
on a 0.5-mm-diameter polyimide tube (The Furukawa Electric Co., Ltd., PIT-S).
The excitation field $H\subm{AC}$ was 8.7~mOe~rms at 887~Hz, which is much lower than the $\Hc$ of \apo.
To reduce the influences of 
remnant magnetic fields such as the earth's field and the residual field in the equipment,
these measurements were performed in a magnetic shield.
We used a cylinder of permalloy (Hamamatsu Photonics K.K., E989-28), 
which has an extremely high permeability.
Inside the permalloy tube, we also placed a lead cylinder with a closed bottom, 
to expel the remaining magnetic flux.
The DC magnetic field for the measurements was applied with a
small solenoidal coil of Nb-Ti superconducting wire placed inside the shield.
The magnitude of the DC field $H\subm{DC}$ is numerically calculated
by taking into account the shielding current on lead shield's surface~\cite{Smith1973,Muething1982}.

\begin{figure}
\includegraphics[width=0.48\textwidth]{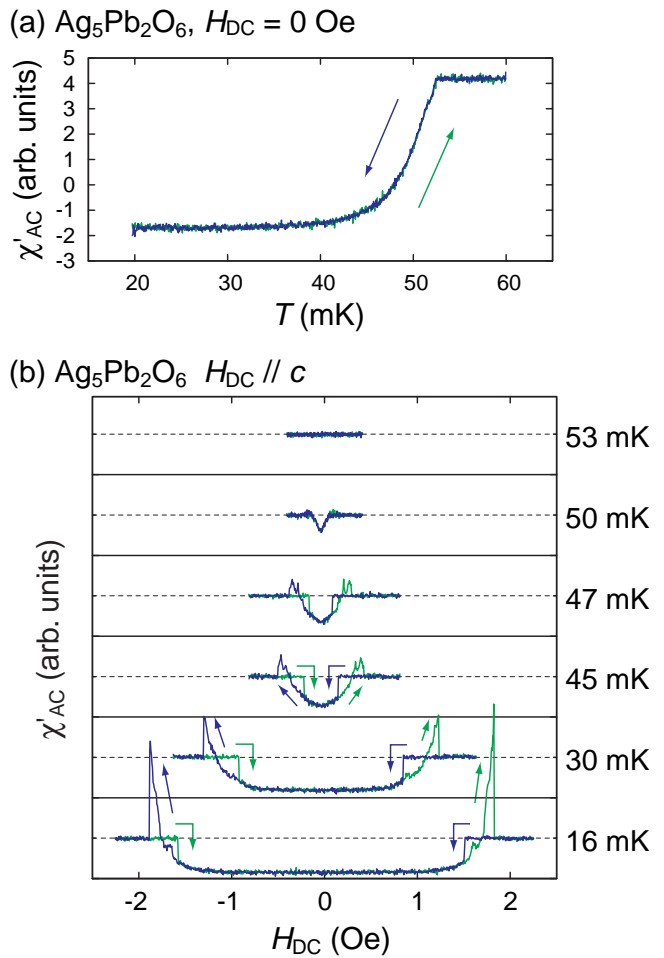}
\caption{(color online) AC susceptibility of \apo.
(a) Result of a temperature sweep with a sweep rate of 0.2~mK/min. 
The residual field $H\subm{res}$ has been compensated in this sweep,
yielding $T\subm{c0}=52.4$~mK.
(b) Results of field sweeps at several temperatures with a sweep rate of 24-47~mOe/min.
From the slight asymmetry of the data, the residual field is estimated as $H\subm{res}=0.040$~Oe.
\label{fig:ac-susceptibility}
}
\end{figure}

The observed zero-resistivity transition is shown in Fig.~\ref{fig:zero-resistivity}.
A clear zero resistivity is seen, which marks definitive evidence of superconductivity of \apo.
We note here that the result in Fig.~\ref{fig:zero-resistivity} 
was obtained without the magnetic shield.
A hysteresis at the superconducting transition and a lower $\Tc$
than that in the AC susceptibility measurement are 
attributable to the influence of the uncanceled residual field.
We confirmed that the hysteresis indeed disappears in the measurement with the magnetic shield.
We next show in Fig.~\ref{fig:ac-susceptibility} 
the real part of the AC susceptibility $\chi\dash\subm{AC}$ of a mono-crystal
with the magnetic shield described below.
It is worth noting that we used the identical crystal for the measurements 
for Figs.~\ref{fig:zero-resistivity} and \ref{fig:ac-susceptibility} 
(see the left inset of Fig.~\ref{fig:zero-resistivity}). 
We also note here that the diamagnetic signal shown in 
Fig.~\ref{fig:ac-susceptibility} is as large as that 
of pure Al with a similar size and shape. 
Such results of the low-frequency susceptibility add a strong 
support for the bulk nature of the superconductivity in \apo.
The measurements were performed under the condition $H\subm{DC}\parallel H\subm{AC} \parallel c$.
The critical temperature $\Tc$ to some extent depends on samples;
the highest $\Tc$ obtained is 52.4~mK, as shown in Fig.~\ref{fig:ac-susceptibility}.

In Fig.~\ref{fig:ac-susceptibility}, 
there are two strong pieces of evidence that \apo\ is a type-I superconductor.
One is the fact that a large supercooling is observed at the superconducting transition under magnetic fields
while no supercooling is seen in zero field.
This means that 
the superconducting transition becomes first order only when an external field is applied.
Such behavior is only seen in type-I superconductors.
The other is the very large positive peaks of $\chi\dash\subm{AC}$ 
just before the superconducting to normal transitions.
These peaks are ascribable to the ``differential paramagnetic effect'' (DPE)~\cite{Hein1961},
which represents that 
the field derivative of the magnetization $\partial M/\partial H$ is positive near the transition and
also the magnetic process in this region is reversible.
In a type-I superconductor with a finite size, the intermediate state
takes place and the DPE should be observed.
On the other hand, type-II superconductors should show no or rather small DPE
because of the irreversibity of magnetic process due to flux pinning.
Thus the large DPE is a hallmark of type-I superconductivity.

\begin{figure}
\includegraphics[width=0.42\textwidth]{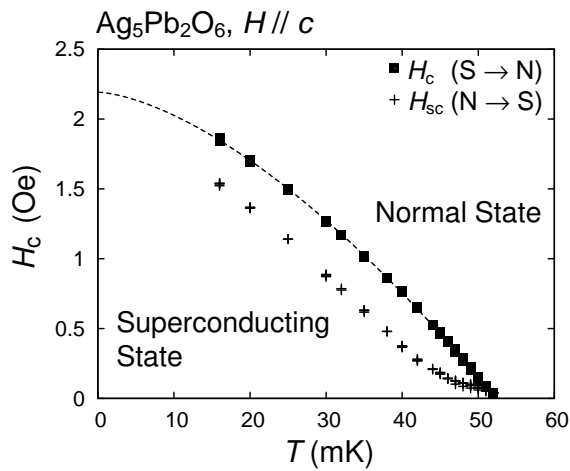}
\caption{Phase diagram of the superconducting phase of \apo,
determined from the field-sweep data of ac susceptibility.
The residual field has been subtracted in the shield $H\subm{res}=0.040$~Oe from the raw data.
The filled squares are superconducting to normal transition field and should be equal to $\Hc$ (see text).
The crosses are the supercooling field $\Hsc$ corresponding to the normal to superconducting transitions.
The broken line is the result of fitting with $\Hc(T)=H\subm{c0}[1-(T/T\subm{c0})^{\alpha}]$.
\label{fig:phase_diagram}
}
\end{figure}

Figure~\ref{fig:phase_diagram} is the phase diagram based on the AC susceptibility of
the crystal with the highest $\Tc$.
Here we identify the transition fields of the superconducting to normal transition as critical fields $\Hc$.
This should be valid despite the possibility of superheating,
because the observed DPE shows that \apo\ is in the intermediate state,
in which superconducting and normal states coexist, 
and there should be no superheating at the ``transition'' from the intermediate state
to the normal state.
We also define the normal to superconducting transition fields as supercooling fields $\Hsc$.
This transition should be from the normal to the full Meissner states since we observed no DPE.
We can fit a relation $\Hc(T)=H\subm{c0}[1-(T/T\subm{c0})^{\alpha}]$ to all the $\Hc$ data down to 16~mK
using $H\subm{c0}$ and $\alpha$ as fitting parameters,
while $T\subm{c0}=52.4$~mK is determined from the temperature sweep data in zero field.
As a result, we obtained $H\subm{c0}=2.19$~Oe and $\alpha=1.56$.
The data can also be fitted by a conventional relation with $\alpha=2$: 
$\Hc(T)=H^\ast\subm{c0}[1-(T/T\subm{c0})^2]$.
However, the fitting is successful only down to $T/\Tc=0.7$ 
and the resulting parameter is $H^\ast\subm{c0}=1.80$~Oe.

Now we can evaluate some of the superconducting parameters from these results.
First, the London penetration depth is obtained as 
$\lambda\subm{L}(0)=(m^{\ast}c^2/4\pi n e^2)^{1/2}=83$~nm.
Here $n=1.0/V\subm{M}=0.51\times 10^{22}\ \mathrm{electrons/cm^3}$ is the electron carrier density, where
$V\subm{M}=0.195\ \mathrm{nm}{}^3$ is the volume of a unit cell~\cite{NoteUnitVolume}, 
and $m^{\ast}=(3\hbar^2 \gamma\subm{e})/(k\subm{B}^2k\subm{F})=1.2m\subm{e}$ is the effective mass.
We used here the measured electronic specific heat coefficient~\cite{Yonezawa2004PRB} 
$\gamma\subm{e}=3.42\ \mathrm{mJ/mol\,K^2} = 291\ \mathrm{erg/cm^3K^2}$,
and the Fermi wavenumber $k\subm{F}=4.5$~nm$^{-1}$ (in the $ab$ plane)~\cite{NoteFermiWavenumber}.
The Ginzburg-Landau (GL) coherence length, 
$\xi_0=(0.18\hbar v\subm{F})/(k\subm{B}\Tc) = 11\ \mathrm{\mu m}$,
where $v\subm{F}=\hbar k\subm{F}/m^{\ast}$ is the Fermi velocity,
is comparable to that of tungsten~\cite{Wheatley1969}
($\xi_0=32\ \mathrm{\mu m},\ \Tc = 15.4\ \mathrm{mK}$).
The mean free path $l$ is given by $l=v\subm{F}\tau$, where $\tau$ is the scattering time of electrons
and has a relation $\tau^{-1}=ne^2\rho/m^\ast$ for the Drude model.
If we use $\rho=1.5\ \mathrm{\mu\Omega\,cm}$, the residual resistivity in the $ab$ plane~\cite{Yonezawa2004PRB},
we obtain $l=240\ \mathrm{nm}$.

One of the important consequences of the above evaluation is that \apo\ is a dirty-limit ($\xi_0\gg l$) superconductor.
This is rather inevitable, since it seems practically impossible to make $l$ longer than $\xi_0$.
In a dirty-limit superconductor,
the GL parameter $\kappa$ is given by 
$\kappa=0.75\lambda\subm{L}(0)/l$ (Ref.~27), and is $0.26$ in our case.
This is indeed smaller than $1/\sqrt{2}$, the border between type-I and -II superconductors,
and is consistent with the type-I behavior of \apo.
The dirty-limit conclusion also implies that the pairing symmetry of the superconductivity is not anisotropic,
because anisotropic superconductivity should be easily suppressed even by non-magnetic impurities.

According to the GL theory, analysis of supercooling gives the upper limit of $\kappa$.
When one decreases the external field of a supercooled superconductor
at constant temperature,
the sample turns into the superconducting state
before the field reaches the ideal supercooling field $H\subm{sc,ideal}$.
If the sample is in vacuum, 
$H\subm{sc,ideal}$ is equal to the surface nucleation field $H\subm{c3}$~\cite{FinkText},
which has a relation $H\subm{c3}=1.695H\subm{c2}=1.695\sqrt{2}\kappa \Hc$.
The observed supercooling field $\Hsc$ satisfies an inequality:
\begin{align}
\Hsc \geq H\subm{sc, ideal} = 1.695\sqrt{2}\kappa \Hc.
\end{align}
Thus $\kappa$ must be smaller than $\kappa\subm{sc}\equiv \Hsc/(1.695\sqrt{2} \Hc)$.

An approach based on this has been used to \textit{determine} $\kappa$ of several pure metals and alloys 
by observing \textit{ideal} supercooling. 
For example, Feder and McLachlan~\cite{Feder1969} realized ideal supercooling of indium and tin 
with precise experiments and obtained
$\kappa\subm{In}=0.0620$ and $\kappa\subm{Sn}=0.0926$.

\begin{figure}
\includegraphics[width=0.45\textwidth]{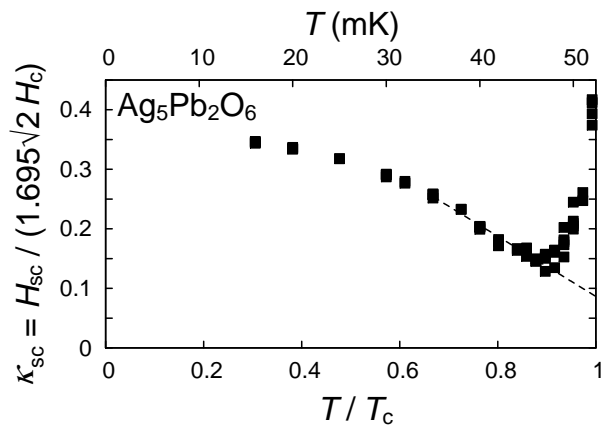}
\caption{The ratio $\kappa\subm{sc}\equiv\Hsc/(1.695\sqrt{2} \Hc)$ of \apo, 
which gives the upper limit of the GL parameter.
The upturn of the graph near $\Tc$ is attributable to the size effect.
The broken line is the result of linear fitting between $T=35$ and 47~mK,
where the size effect is not significant.
\label{fig:kappa-T}
}
\end{figure}

We calculated $\kappa\subm{sc}$ of \apo\ at each temperature as shown in Fig.~\ref{fig:kappa-T}.
The steep increase of $\kappa\subm{sc}$ close to $\Tc$ is attributed to the size effect,
which occurs when the temperature-dependent coherence length $\xi(T)\propto[\Tc/(\Tc-T)]^{1/2}\xi$ 
becomes comparable to the size of a sample.
In fact, the coherence length,
being $\xi\sim(\xi_0 l)^{1/2}$ in a dirty-limit superconductor~\cite{deGennesTextChap7}, 
becomes $1.4\ \mathrm{\mu m}$.
This is large enough to cause the size effect near $\Tc$ in a sample of 100-200~$\mu$m ($\sim 100\xi$).
Indeed, in the experiments of Feder and McLachlan~\cite{Feder1969} a sphere of clean indium 
($\xi\sim\xi_0=0.20\ \mathrm{\mu m}$)
with a radius 16~$\mu$m ($\sim 80\xi$) showed the size effect near $\Tc$.

Feder and McLachlan determined $\kappa$ by extrapolating $\kappa(T)$ to $T=\Tc$,
because the influence of nucleation centers becomes negligible near $\Tc$
due to divergence of $\xi(T)$.
Following their procedure we extrapolated $\kappa\subm{sc}(T)$ in $35\ \mathrm{mK}<T<47\ \mathrm{mK}$ to $\Tc$ as the broken line in Fig.~\ref{fig:kappa-T}.
The extrapolation gives $\kappa\subm{sc}|_{T=\Tc}=0.085$,
which should be the upper limit of $\kappa$ of \apo.
This estimated upper limit is smaller than the calculated value but
the difference should be within a consistency.
In the case of indium or tin~\cite{Feder1969}, 
there are also differences of a few factors among values of $\kappa$ obtained by different procedures.

In conclusion, we succeeded in observing zero resistivity transition of a silver oxide \apo,
giving definitive evidence of the long-awaited layered silver oxide superconductor
since the discovery of high-$\Tc$ cuprates.
The AC susceptibility reveals that \apo\ is an oxide type-I superconductor.
It is widely considered that type-I superconductivity is rare in compounds,
although there is no fundamental reason to prohibit it.
The present discovery indeed demonstrates that even an oxide can be an extreme type-I superconductor.
Superconducting parameters indicate 
that \apo\ is a dirty-limit type-I superconductor and 
thus the pairing symmetry of \apo\ should be isotropic.
The discovery of this novel class of superconductor, 
silver oxide superconductor with a layered structure, should motivate 
searches for more superconductors among similar silver oxides.
A salient next target is to adjust the doping to realize the electronic states with strong electron correlations,
closely analogous to that of the high-$\Tc$ cuprates,
in order to seek for unconventional superconductivity.

\begin{acknowledgements}
We would like to acknowledge K.~Ishida, S.~Nakatsuji, H.~Yaguchi,
K.~Deguchi and K.~Kitagawa for their support,
K.~Yamada, R.~Ikeda and S.~Fujimoto for helpful discussions,
and Y.~Sasaki for his advise on experimental techniques.
We also appreciate Furukawa Electric Co., Ltd. for providing us with polyimide tubes.
This work has been supported by a Grant-in-Aid for the 21st Century COE 
``Center for Diversity and Universality in Physics'' from  Ministry of Education, Culture, Sports, Science and 
Technology (MEXT) of Japan.
It has also been supported by Grants-in-Aids for Scientific Research from MEXT 
and from Japan Society for the Promotion of Science (JSPS).
\end{acknowledgements}

\bibliography{Ag5Pb2O6_2005}

\begin{thebibliography}{27}
\expandafter\ifx\csname natexlab\endcsname\relax\def\natexlab#1{#1}\fi
\expandafter\ifx\csname bibnamefont\endcsname\relax
  \def\bibnamefont#1{#1}\fi
\expandafter\ifx\csname bibfnamefont\endcsname\relax
  \def\bibfnamefont#1{#1}\fi
\expandafter\ifx\csname citenamefont\endcsname\relax
  \def\citenamefont#1{#1}\fi
\expandafter\ifx\csname url\endcsname\relax
  \def\url#1{\texttt{#1}}\fi
\expandafter\ifx\csname urlprefix\endcsname\relax\def\urlprefix{URL }\fi
\providecommand{\bibinfo}[2]{#2}
\providecommand{\eprint}[2][]{\url{#2}}

\bibitem[{\citenamefont{Nagata and Atake}(1999)}]{Nagata1999}
\bibinfo{author}{\bibfnamefont{S.}~\bibnamefont{Nagata}} \bibnamefont{and}
  \bibinfo{author}{\bibfnamefont{T.}~\bibnamefont{Atake}}, \bibinfo{journal}{J.
  Therm. Anal. Calorim.} \textbf{\bibinfo{volume}{57}}, \bibinfo{pages}{807}
  (\bibinfo{year}{1999}).

\bibitem[{\citenamefont{Cava}(2000)}]{Cava2000}
\bibinfo{author}{\bibfnamefont{R.~J.} \bibnamefont{Cava}}, \bibinfo{journal}{J.
  Am. Ceram. Soc.} \textbf{\bibinfo{volume}{83}}, \bibinfo{pages}{5}
  (\bibinfo{year}{2000}).

\bibitem[{\citenamefont{Bednorz and M{\"u}ller}(1986)}]{Bednorz1986}
\bibinfo{author}{\bibfnamefont{J.~G.} \bibnamefont{Bednorz}} \bibnamefont{and}
  \bibinfo{author}{\bibfnamefont{K.~A.} \bibnamefont{M{\"u}ller}},
  \bibinfo{journal}{Z. Phys. B} \textbf{\bibinfo{volume}{64}},
  \bibinfo{pages}{189} (\bibinfo{year}{1986}).

\bibitem[{\citenamefont{Maeno et~al.}(1994)\citenamefont{Maeno, Hashimoto,
  Yoshida, Nishizaki, Fujita, Bednorz, and Lichtenberg}}]{Maeno1994Nature}
\bibinfo{author}{\bibfnamefont{Y.}~\bibnamefont{Maeno}},
  \bibinfo{author}{\bibfnamefont{H.}~\bibnamefont{Hashimoto}},
  \bibinfo{author}{\bibfnamefont{K.}~\bibnamefont{Yoshida}},
  \bibinfo{author}{\bibfnamefont{S.}~\bibnamefont{Nishizaki}},
  \bibinfo{author}{\bibfnamefont{T.}~\bibnamefont{Fujita}},
  \bibinfo{author}{\bibfnamefont{J.~G.} \bibnamefont{Bednorz}},
  \bibnamefont{and}
  \bibinfo{author}{\bibfnamefont{F.}~\bibnamefont{Lichtenberg}},
  \bibinfo{journal}{Nature (London)} \textbf{\bibinfo{volume}{372}},
  \bibinfo{pages}{532} (\bibinfo{year}{1994}).

\bibitem[{\citenamefont{Mackenzie and Maeno}(2003)}]{Mackenzie2003RMP}
\bibinfo{author}{\bibfnamefont{A.~P.} \bibnamefont{Mackenzie}}
  \bibnamefont{and} \bibinfo{author}{\bibfnamefont{Y.}~\bibnamefont{Maeno}},
  \bibinfo{journal}{Rev. Mod. Phys.} \textbf{\bibinfo{volume}{75}},
  \bibinfo{pages}{657} (\bibinfo{year}{2003}).

\bibitem[{\citenamefont{Takada et~al.}(2003)\citenamefont{Takada, Sakurai,
  Takayama-Muromachi, Izumi, Dilanian, and Sasaki}}]{Takada2003}
\bibinfo{author}{\bibfnamefont{K.}~\bibnamefont{Takada}},
  \bibinfo{author}{\bibfnamefont{H.}~\bibnamefont{Sakurai}},
  \bibinfo{author}{\bibfnamefont{E.}~\bibnamefont{Takayama-Muromachi}},
  \bibinfo{author}{\bibfnamefont{F.}~\bibnamefont{Izumi}},
  \bibinfo{author}{\bibfnamefont{R.~A.} \bibnamefont{Dilanian}},
  \bibnamefont{and} \bibinfo{author}{\bibfnamefont{T.}~\bibnamefont{Sasaki}},
  \bibinfo{journal}{Nature (London)} \textbf{\bibinfo{volume}{422}},
  \bibinfo{pages}{53} (\bibinfo{year}{2003}).

\bibitem[{\citenamefont{Robin et~al.}(1966)\citenamefont{Robin, Andres,
  Geballe, Kuebler, and McWhan}}]{Robin1966}
\bibinfo{author}{\bibfnamefont{M.~B.} \bibnamefont{Robin}},
  \bibinfo{author}{\bibfnamefont{K.}~\bibnamefont{Andres}},
  \bibinfo{author}{\bibfnamefont{T.~H.} \bibnamefont{Geballe}},
  \bibinfo{author}{\bibfnamefont{N.~A.} \bibnamefont{Kuebler}},
  \bibnamefont{and} \bibinfo{author}{\bibfnamefont{D.~B.}
  \bibnamefont{McWhan}}, \bibinfo{journal}{Phys. Rev. Lett.}
  \textbf{\bibinfo{volume}{17}}, \bibinfo{pages}{917} (\bibinfo{year}{1966}).

\bibitem[{\citenamefont{Tsindlekht et~al.}(2004)\citenamefont{Tsindlekht,
  Leviev, Asulin, Sharoni, Millo, Felner, Paderno, Filippov, and
  Belogolovskii}}]{Tsindlekht2004}
\bibinfo{author}{\bibfnamefont{M.~I.} \bibnamefont{Tsindlekht}},
  \bibinfo{author}{\bibfnamefont{G.~I.} \bibnamefont{Leviev}},
  \bibinfo{author}{\bibfnamefont{I.}~\bibnamefont{Asulin}},
  \bibinfo{author}{\bibfnamefont{A.}~\bibnamefont{Sharoni}},
  \bibinfo{author}{\bibfnamefont{O.}~\bibnamefont{Millo}},
  \bibinfo{author}{\bibfnamefont{I.}~\bibnamefont{Felner}},
  \bibinfo{author}{\bibfnamefont{Y.~B.} \bibnamefont{Paderno}},
  \bibinfo{author}{\bibfnamefont{V.~B.} \bibnamefont{Filippov}},
  \bibnamefont{and} \bibinfo{author}{\bibfnamefont{M.~A.}
  \bibnamefont{Belogolovskii}}, \bibinfo{journal}{Phys. Rev. B}
  \textbf{\bibinfo{volume}{69}}, \bibinfo{pages}{212508}
  (\bibinfo{year}{2004}).

\bibitem[{\citenamefont{Yamaguchi et~al.}(1987)\citenamefont{Yamaguchi, Waki,
  and Mitsugi}}]{Yamaguchi1987}
\bibinfo{author}{\bibfnamefont{Y.}~\bibnamefont{Yamaguchi}},
  \bibinfo{author}{\bibfnamefont{S.}~\bibnamefont{Waki}}, \bibnamefont{and}
  \bibinfo{author}{\bibfnamefont{K.}~\bibnamefont{Mitsugi}},
  \bibinfo{journal}{J. Phys. Soc. Jpn.} \textbf{\bibinfo{volume}{56}},
  \bibinfo{pages}{419} (\bibinfo{year}{1987}).

\bibitem[{\citenamefont{Hull et~al.}(1981)\citenamefont{Hull, Wernick, Geballe,
  Waszczak, and Bernardini}}]{Hull1981}
\bibinfo{author}{\bibfnamefont{G.~W.} \bibnamefont{Hull}},
  \bibinfo{author}{\bibfnamefont{J.~H.} \bibnamefont{Wernick}},
  \bibinfo{author}{\bibfnamefont{T.~H.} \bibnamefont{Geballe}},
  \bibinfo{author}{\bibfnamefont{J.~V.} \bibnamefont{Waszczak}},
  \bibnamefont{and} \bibinfo{author}{\bibfnamefont{J.~E.}
  \bibnamefont{Bernardini}}, \bibinfo{journal}{Phys. Rev. B}
  \textbf{\bibinfo{volume}{24}}, \bibinfo{pages}{6715} (\bibinfo{year}{1981}).

\bibitem[{\citenamefont{Palstra et~al.}(1986)\citenamefont{Palstra, Lu,
  Menovsky, Nieuwenhuys, Kes, and Mydosh}}]{Palstra1986}
\bibinfo{author}{\bibfnamefont{T.~T.~M.} \bibnamefont{Palstra}},
  \bibinfo{author}{\bibfnamefont{G.}~\bibnamefont{Lu}},
  \bibinfo{author}{\bibfnamefont{A.~A.} \bibnamefont{Menovsky}},
  \bibinfo{author}{\bibfnamefont{G.~J.} \bibnamefont{Nieuwenhuys}},
  \bibinfo{author}{\bibfnamefont{P.~H.} \bibnamefont{Kes}}, \bibnamefont{and}
  \bibinfo{author}{\bibfnamefont{J.~A.} \bibnamefont{Mydosh}},
  \bibinfo{journal}{Phys. Rev. B} \textbf{\bibinfo{volume}{34}},
  \bibinfo{pages}{4566} (\bibinfo{year}{1986}).

\bibitem[{\citenamefont{Gottlieb et~al.}(1986)\citenamefont{Gottlieb,
  Lasjaunias, Tholence, Laborde, Thomas, and Madar}}]{Gottlieb1992}
\bibinfo{author}{\bibfnamefont{U.}~\bibnamefont{Gottlieb}},
  \bibinfo{author}{\bibfnamefont{J.~C.} \bibnamefont{Lasjaunias}},
  \bibinfo{author}{\bibfnamefont{J.~L.} \bibnamefont{Tholence}},
  \bibinfo{author}{\bibfnamefont{O.}~\bibnamefont{Laborde}},
  \bibinfo{author}{\bibfnamefont{O.}~\bibnamefont{Thomas}}, \bibnamefont{and}
  \bibinfo{author}{\bibfnamefont{R.}~\bibnamefont{Madar}},
  \bibinfo{journal}{Phys. Rev. B} \textbf{\bibinfo{volume}{45}},
  \bibinfo{pages}{4803} (\bibinfo{year}{1986}).

\bibitem[{\citenamefont{Soulen et~al.}(1977)\citenamefont{Soulen, Utton, and
  Colwell}}]{Soulen1977}
\bibinfo{author}{\bibfnamefont{R.~J.} \bibnamefont{Soulen}},
  \bibinfo{author}{\bibfnamefont{D.~B.} \bibnamefont{Utton}}, \bibnamefont{and}
  \bibinfo{author}{\bibfnamefont{J.~H.} \bibnamefont{Colwell}},
  \bibinfo{journal}{Bull. Amer. Phys. Soc.} \textbf{\bibinfo{volume}{22}},
  \bibinfo{pages}{403} (\bibinfo{year}{1977}).

\bibitem[{\citenamefont{Kobayashi and Tsujikawa}(1981)}]{Kobayashi1981}
\bibinfo{author}{\bibfnamefont{M.}~\bibnamefont{Kobayashi}} \bibnamefont{and}
  \bibinfo{author}{\bibfnamefont{I.}~\bibnamefont{Tsujikawa}},
  \bibinfo{journal}{J. Phys. Soc. Jpn.} \textbf{\bibinfo{volume}{50}},
  \bibinfo{pages}{3245} (\bibinfo{year}{1981}).

\bibitem[{\citenamefont{Bystr{\"{o}}m and Evers}(1950)}]{Bystrom1950}
\bibinfo{author}{\bibfnamefont{A.}~\bibnamefont{Bystr{\"{o}}m}}
  \bibnamefont{and} \bibinfo{author}{\bibfnamefont{L.}~\bibnamefont{Evers}},
  \bibinfo{journal}{Acta Chem. Scand.} \textbf{\bibinfo{volume}{4}},
  \bibinfo{pages}{613} (\bibinfo{year}{1950}).

\bibitem[{\citenamefont{Jansen et~al.}(1990)\citenamefont{Jansen, Bortz, and
  Heidebrecht}}]{Jansen1990}
\bibinfo{author}{\bibfnamefont{M.}~\bibnamefont{Jansen}},
  \bibinfo{author}{\bibfnamefont{M.}~\bibnamefont{Bortz}}, \bibnamefont{and}
  \bibinfo{author}{\bibfnamefont{K.}~\bibnamefont{Heidebrecht}},
  \bibinfo{journal}{J. Less-Common Met.} \textbf{\bibinfo{volume}{161}},
  \bibinfo{pages}{17} (\bibinfo{year}{1990}).

\bibitem[{\citenamefont{Brennan and Burdett}(1994)}]{Brennan1993}
\bibinfo{author}{\bibfnamefont{T.~D.} \bibnamefont{Brennan}} \bibnamefont{and}
  \bibinfo{author}{\bibfnamefont{J.~K.} \bibnamefont{Burdett}},
  \bibinfo{journal}{Inorg. Chem.} \textbf{\bibinfo{volume}{33}},
  \bibinfo{pages}{4794} (\bibinfo{year}{1994}).

\bibitem[{\citenamefont{Yonezawa and Maeno}(2004)}]{Yonezawa2004PRB}
\bibinfo{author}{\bibfnamefont{S.}~\bibnamefont{Yonezawa}} \bibnamefont{and}
  \bibinfo{author}{\bibfnamefont{Y.}~\bibnamefont{Maeno}},
  \bibinfo{journal}{Phys. Rev. B} \textbf{\bibinfo{volume}{70}},
  \bibinfo{pages}{184523} (\bibinfo{year}{2004}).

\bibitem[{\citenamefont{Smith}(1973)}]{Smith1973}
\bibinfo{author}{\bibfnamefont{T.~I.} \bibnamefont{Smith}},
  \bibinfo{journal}{Rev. Sci. Instrum.} \textbf{\bibinfo{volume}{44}},
  \bibinfo{pages}{852} (\bibinfo{year}{1973}).

\bibitem[{\citenamefont{Muething et~al.}(1982)\citenamefont{Muething, Edwards,
  Feder, Gully, and Scholz}}]{Muething1982}
\bibinfo{author}{\bibfnamefont{K.~A.} \bibnamefont{Muething}},
  \bibinfo{author}{\bibfnamefont{D.~O.} \bibnamefont{Edwards}},
  \bibinfo{author}{\bibfnamefont{J.~D.} \bibnamefont{Feder}},
  \bibinfo{author}{\bibfnamefont{W.~J.} \bibnamefont{Gully}}, \bibnamefont{and}
  \bibinfo{author}{\bibfnamefont{H.~N.} \bibnamefont{Scholz}},
  \bibinfo{journal}{Rev. Sci. Instrum.} \textbf{\bibinfo{volume}{53}},
  \bibinfo{pages}{485} (\bibinfo{year}{1982}).

\bibitem[{\citenamefont{Hein and Falge}(1961)}]{Hein1961}
\bibinfo{author}{\bibfnamefont{R.~A.} \bibnamefont{Hein}} \bibnamefont{and}
  \bibinfo{author}{\bibfnamefont{R.~L.} \bibnamefont{Falge},
  \bibfnamefont{Jr}}, \bibinfo{journal}{Phys. Rev.}
  \textbf{\bibinfo{volume}{123}}, \bibinfo{pages}{407} (\bibinfo{year}{1961}).

\bibitem[{Not({\natexlab{a}})}]{NoteUnitVolume}
\bibinfo{note}{The value of {$V_{\mathrm{M}}$}, which is a room temperature
  value, is taken from {Ref.~16}. Stricktly speaking, we should use the value
  at {$T_{\mathrm{c}}$}. However, termperature variation of {$V_{\mathrm{M}}$}
  should be less than a few percent.}

\bibitem[{Not({\natexlab{b}})}]{NoteFermiWavenumber}
\bibinfo{note}{The value of {$k_{\mathrm{F}}$} is quated from {Ref.~17}. Since
  the figure is not given in that article, we estimated its value from the
  graph of band dispersion.}

\bibitem[{\citenamefont{Wheatley et~al.}(1969)\citenamefont{Wheatley, Johnson,
  and Black}}]{Wheatley1969}
\bibinfo{author}{\bibfnamefont{J.~C.} \bibnamefont{Wheatley}},
  \bibinfo{author}{\bibfnamefont{R.~T.} \bibnamefont{Johnson}},
  \bibnamefont{and} \bibinfo{author}{\bibfnamefont{W.~C.} \bibnamefont{Black}},
  \bibinfo{journal}{J. Low Temp. Phys.} \textbf{\bibinfo{volume}{1}},
  \bibinfo{pages}{641} (\bibinfo{year}{1969}).

\bibitem[{\citenamefont{Fink et~al.}(1978)\citenamefont{Fink, McLachlan, and
  Rothberg-Bibby}}]{FinkText}
\bibinfo{author}{\bibfnamefont{H.~J.} \bibnamefont{Fink}},
  \bibinfo{author}{\bibfnamefont{D.~S.} \bibnamefont{McLachlan}},
  \bibnamefont{and}
  \bibinfo{author}{\bibfnamefont{B.}~\bibnamefont{Rothberg-Bibby}}, in
  \emph{\bibinfo{booktitle}{Progress in {L}ow {T}emperature {P}hysics}}, edited
  by \bibinfo{editor}{\bibfnamefont{D.~F.} \bibnamefont{Brewer}}
  (\bibinfo{publisher}{North-Holland, Amsterdam}, \bibinfo{year}{1978}), vol.
  \bibinfo{volume}{VII B}, chap.~\bibinfo{chapter}{6}.

\bibitem[{\citenamefont{Feder and McLachlan}(1969)}]{Feder1969}
\bibinfo{author}{\bibfnamefont{J.}~\bibnamefont{Feder}} \bibnamefont{and}
  \bibinfo{author}{\bibfnamefont{D.~S.} \bibnamefont{McLachlan}},
  \bibinfo{journal}{Phys. Rev.} \textbf{\bibinfo{volume}{177}},
  \bibinfo{pages}{763} (\bibinfo{year}{1969}).

\bibitem[{\citenamefont{de~Gennes}(1966)}]{deGennesTextChap7}
\bibinfo{author}{\bibfnamefont{P.~G.} \bibnamefont{de~Gennes}},
  \emph{\bibinfo{title}{Superconductivity of metals and alloys}}
  (\bibinfo{publisher}{W. A. Benjamin, New York}, \bibinfo{year}{1966}).

\end{thebibliography}

\end{document}